\documentstyle[10pt,epsfig,supertabular]{elsart}
\input{epsf.sty}

\newcommand{\Laena}{\Lambda_{\mathrm Ae}N_{\mathrm A}}

\newcommand{\Ldnd}{\Lambda_{\mathrm D}N_{\mathrm D}}

\begin{document}
%\tableofcontents
\begin{frontmatter}
\title {Response of CsI(Tl) scintillators over a large range in energy 
and atomic number of ions (Part II): calibration and identification in
the INDRA array}
\small
\author[buch,ipno]{M.~P\^arlog},
\author[ipno]{B.~Borderie\thanksref{cores}}, 
\author[ipno]{M.F.~Rivet},
\author[buch,ipno]{G.~T\u{a}b\u{a}caru},  
\author[ganil]{A.~Chbihi},
\author[maroc]{M. Elouardi},
\author[lpc]{N.~Le Neindre},
\author[lpc]{O.~Lopez}, 
\author[ipno]{E.~Plagnol}, 
\author[ipno]{L.~Tassan-Got},
\author[ganil]{G.~Auger},
\author[ipno]{Ch.O.~Bacri}, 
\author[lpc]{N.~Bellaize}, 
\author[lpc]{F.~Bocage},
\author[lpc]{R.~Bougault}, 
\author[ganil]{B.~Bouriquet}, 
\author[lpc]{R.~Brou}, 
\author[cea]{P.~Buchet},
\author[cea]{J.L.~Charvet}, 
\author[lpc]{J.~Colin}, 
\author[lpc]{D.~Cussol}, 
\author[cea]{R.~Dayras},
\author[ipnl]{A.~Demeyer}, 
\author[cea]{D.~Dor\'e}, 
\author[lpc]{D.~Durand}, 
\author[ganil]{J.D.~Frankland},
\author[ipno,cnam]{E.~Galichet}, 
\author[lpc]{E.~Genouin-Duhamel}, 
\author[ipnl]{E.~Gerlic},
\author[ganil]{S.~Hudan}, 
\author[ipnl]{D.~Guinet}, 
\author[ipnl]{P.~Lautesse}, 
\author[ipno]{F.~Lavaud},
\author[ganil]{J.L.~Laville}, 
\author[lpc]{J.F.~Lecolley}, 
\author[ipnl]{C.~Leduc}, 
\author[cea]{R.~Legrain},
\author[lpc]{M.~Louvel}, 
\author[ipnl]{A.M.~Maskay}, 
\author[cea]{L.~Nalpas}, 
\author[lpc]{J.~Normand},
\author[lpc]{J.~P\'{e}ter}, 
\author[napol]{E.~Rosato}, 
\author[ganil]{F.~Saint-Laurent\thanksref{a}},
\author[lpc]{J.C.~Steckmeyer}, 
%\author[ipnl]{M.~Stern}, 
\author[lpc]{B.~Tamain}, 
\author[ganil]{O.~Tirel},
\author[lpc]{E.~Vient}, 
\author[cea]{C.~Volant}, 
\author[ganil]{J.P.~Wieleczko}

\collaboration{(INDRA collaboration)}
\address[buch]{National Institute for Physics and Nuclear Engineering,
RO-76900 Bucharest-M\u{a}gurele, Romania}
\address[ipno]{Institut de Physique Nucl\'eaire, IN2P3-CNRS, F-91406 Orsay
 Cedex, France.}
\address[ganil]{GANIL, CEA et IN2P3-CNRS, B.P.~5027, F-14076 Caen Cedex, France.}
\address[maroc]{Laboratoire de Physique Nucl\'eaire Appliqu\'ee,
Kenitra, Maroc.}
\address[lpc]{LPC, IN2P3-CNRS, ISMRA et Universit\'e, F-14050 Caen Cedex,
France.}
\address[cea]{DAPNIA/SPhN, CEA/Saclay, F-91191 Gif sur Yvette Cedex,
France.}
\address[ipnl]{Institut de Physique Nucl\'eaire, IN2P3-CNRS et Universit\'e,
F-69622 Villeurbanne
Cedex, France.}
\address[napol]{Dipartimento di Scienze Fisiche e Sezione INFN, Universit\`a
di Napoli ``Federico II'', I80126 Napoli, Italy.}
\address[cnam]{Conservatoire National des Arts et M\'etiers, F-75141
Paris cedex 03.}
\thanks[cores]{Corresponding author. Tel 33 1 69157148; fax 33 1
69154507; e-mail borderie@ ipno.in2p3.fr} 
\thanks[a]{present address: DRFC/STEP, CEA/Cadarache, F-13018
Saint-Paul-lez-Durance Cedex, France.}
%\thanks[pres-ipnl]{present address: Institut de Physique Nucl\'eaire, IN2P3-CNRS et Universit\'e,
%F-69622 Villeurbanne
%Cedex, France.}

%\address{(INDRA collaboration) \\
%$^1$ GANIL, CEA et IN2P3-CNRS, B.P.~5027, F-14076 Caen Cedex, France.~\\
%$^2$ Institut de Physique Nucl\'eaire, IN2P3-CNRS, F-91406 Orsay Cedex,
%France.~\\
%$^3$ LPC, IN2P3-CNRS, ISMRA et Universit\'e, F-14050 Caen Cedex, France.~\\
%$^4$ DAPNIA/SPhN, CEA/Saclay, F-91191 Gif sur Yvette Cedex, France.~\\
%$^5$ Institut de Physique Nucl\'eaire, IN2P3-CNRS et Universit\'e, 
%F-69622 Villeurbanne Cedex, France. \\
%$^6$ National Institute for Physics and Nuclear Engineering, RO-76900 
%Bucharest-M\u{a}gurele, Romania.~\\
%$^7$ Dipartimento di Scienze Fisiche e Sezione INFN, Universit\`a di Napoli
%``Federico II'', I-80126 Napoli, Italy.~\\
%$^8$ Conservatoire National des Arts et M\'etiers, F-75141 Paris cedex 03. \\ 
%$^9$ Laboratoire de Physique Nucl\'eaire Appliqu\'ee, Kenitra, Maroc.\\
%a) present address: DRFC/STEP, CEA/Cadarache, F-13018
%Saint-Paul-lez-Durance Cedex, France
%}
\begin{abstract}
%\newpage
The light output of the 324 CsI(Tl) scintillators of INDRA has been measured
over a large range both in energy: $1 - 80$ AMeV and in atomic number of
incident ions: $Z = 1 - 60$. An analytical expression for the nonlinear total
light response as a function of the energy and the identity of the ion is
proposed. It depends on four parameters. For three of them, connected
to CsI(Tl) intrinsic characteristics, recommended fixed values are proposed.
They are issued from the comparative study of the forward scintillators of
INDRA, based on intermediate mass fragment data. The fourth one, related
to light collection and to the gain of the associated photomultiplier, is
particular and may be accurately obtained afterwards, from light charged
particle data. Two applications are presented: fragment identification in  
%$\Delta E - Light_{CsI(Tl)}$ telescope-type map 
telescopes using a CsI(Tl) crystal as residual energy detector and the
scintillator energy calibration. The results are successfully confronted to 
heavy fragment experimental data.
\end{abstract}
\end{frontmatter}
%%%%%%%%%%%%%%%%%%%%%%%%%%%%%%%%%%%%%%%%%%%%%%%%%%%%%%%%%%%%%%%%%%%%%
%%%%%%%%%%%%%%%%%%%%%%%%%%%%%%%%%%%%%%%%%%%%%%%%%%%%%%%%%%%%%%%%%%%%%
%\tableofcontents
%%%%%%%%%%%%%%%%%%%%%%%%%%%%%%%%%%%%%%%%%%%%%%%%%%%%%%%%%%%%%%%%%%%%%
%%%%%%%%%%%%%%%%%%%%%%%%%%%%%%%%%%%%%%%%%%%%%%%%%%%%%%%%%%%%%%%%%%%%%
\section{Introduction\label{sect1}}
INDRA is a 4$\pi$ axially symmetrical array for the detection of light and
heavy charged nuclear reaction products \cite{pou95,pou96} covering huge
dynamic ranges, both in energy (from $\approx$1 MeV to the maximum available
energy $\approx$6 GeV at GANIL) and in atomic number (from proton to
uranium). It has a high granularity and a shell structure, consisting
of several detection layers. For the last layer, which should stop all particles
and fragments produced using GANIL beams, thallium-activated caesium iodide
scintillators (CsI(Tl)) coupled to photomultiplier tubes were chosen.
\par
The standard calibration procedure for CsI(Tl) is to find a function $L$
depending on the energy, but also on the identity of the particle, which
describes reasonably well the induced scintillation $Q_0$. The parameters of 
this function are determined by a global fit procedure which simultaneousely 
compares the calculated scintillator response to the experimental one for all 
particles and fragments of well known energies. Afterwards, the unknown 
energies of the reaction products detected in physical runs are found by means 
of the calibration function and the related parameters, starting from the 
measured associated light outputs.
Once this is accomplished, one may get a reference map for the identification 
of the reaction products in a two-dimensional plot showing the energy deposited
in the preceding detection layer, for every fragment punching through, versus 
the total light from the CsI(Tl) crystal. This is the second aim of our work.
\par
For the forward angles of INDRA (3$^\circ$ $<$ $\theta$ $<$ 45$^\circ$, rings 
2 - 9), the detection layers which precede the scintillators consists of gas
ionisation chambers (ICs) and 300 $\mu$m silicon detectors. The Si detectors  
have allowed an accurate determination of the residual energy - as presented
in subsection \ref{subsect22} - for the whole range of fragments passing through
and stopped in the scintillators. Thus, there is a tremendous set of data 
which has facilitated a detailed study of the CsI(Tl) crystal light response, 
having led to a convenient expression which describes the total scintillation. 
\par
Both components of the procedure, measured scintillation and its modelling
function, have implied special processing.
In the case of INDRA, neither the amplitude of the CsI(Tl) associated signals
nor the integral of these signals are measured. Instead, fractions of the
total signal are integrated into two time gates, allowing particle
identification. Both integrals have rather complex dependences versus the
incident particle energy as compared to the whole integral. It is possible
to accurately find the latter quantity by software, starting from the two
measured signal fractions, as shown in %subsection \ref{subsect23}.
sections \ref{sect2} and \ref{sect3}.
\par
The exact expression of the total light output of a CsI(Tl) crystal, as
predicted by the recombination and nuclear quenching model (RNQM)
\cite{par00i}, implies a numerical integration over the energy and this
fact is prohibitive for application purposes. Under suitable approximations,
the integration may be however analytically performed and a very easily to
handle light response expression is deduced. Starting from intermediate mass
fragments of known energies, the 3 or 4 involved parameters are determined.
Except for the gain parameter, proper to each scintillator crystal and
corresponding electronic chain, the values of the other associated fit
parameters are fixed, as being connected to intrinsic CsI(Tl) crystal
properties. Procedures for fragment identification in a
%$\Delta E - Light_{CsI(Tl)}$
$\Delta E - Q_0$ telescope-type map, with the scintillator as 
residual energy detector, as well as for the energy calibration of the latter 
one - in the whole dynamic range - are developed and critically analyzed in 
section \ref{sect4}, containing the RNQM applications. 
\par
For the backward angles of INDRA (45$^\circ$ $<$ $\theta$ $<$ 176$^\circ$,
rings 10 - 17), the scintillators are preceded only by ionization chambers.
The calibration of the scintillators leans on the above mentioned light
response expression and the CsI(Tl)
characteristic parameter values, found at forward angles; the individual gain
parameter is determined by means of light charged particle and eventually
light fragment data. The calibration so found allows to safely extrapolate the
charge identification in a %$\Delta E - Light_{CsI(Tl)}$
$\Delta E - Q_0$ map to regions where no 
ridge lines are visible because of very low statistics, improving then both 
charge and energy determination for heavy fragments ($Z \geq 15$) detected 
beyond $45^\circ$ with INDRA. Details are given in section \ref{sect4}.
\par
Our findings are summarized in section \ref{sect5}.
\par
%\newpage						%newpage
Notation and values of physical constants and variables used in this paper.
See also those in the preceding paper \cite{par00i}.
\newpage
%\par
{\small
%\begin{tabular}{cll}
%\begin{tabular}{cl}
\tablefirsthead{
 \hline
Symbol$^*$ & Definition & Units or Value \\
 \hline}
\tablehead{
 \hline
Symbol$^*$ & Definition & Units or Value \\
 \hline
 }
\tabletail{
 \hline
 }
 \par
\begin{supertabular}{cp{7cm}p{4cm}}
%	&	&	\\
\multicolumn{3}{c}{\it Experimental light output and related variables} \\
%	&	&	\\
%$t, dt$ & time, infinitesimal time element respectively & s \\
$t$ & time & s \\
$i(t)$ & signal at the last dynode of the CsI(Tl) photomultiplier (PMT) &
a.u.~s$^{-1}$ \\
$Q_f$ & integral charge of the fast component of the signal & a.u. \\
$\tau_f$ & decay time constant of the fast component & s \\
$Q_s$ & integral charge of the slow component of the signal & a.u. \\
$\tau_s$ & decay time constant of the slow component  & s \\
$Q_{fs}$ & integral charge corresponding to the whole signal & a.u. \\
%$dq(t)$ infinitesimal measured charge at the CsI(Tl) photomultiplier output & a.u. \\
$i_{mes}(t)$ & measured signal at the output of the PMT anodic circuit &
a.u.~s$^{-1}$ \\
%$V$ & measured signal visualized at oscilloscope & V \\
$Q_0$ & approximate total integrated charge $\propto$ experimental light output & a.u. \\
$\tau_0$ & decay time constant & s \\
$\tau_{0min}$ & lower value of the decay time constant & s \\
%$\tau$=RC & time constant of the PMT anodic circuit & s \\
$\tau$ & rise time constant at output of the PMT & s \\
$F$ & experimental charge integrated in the ``fast" gate & a.u. \\
$S$ & experimental charge integrated in the ``slow" gate & a.u. \\
%$R$ & resistance of the PMT anodic circuit & $\Omega$ \\
%$C$ & capacity of the PMT anodic circuit & F \\
%	&	&	\\
\multicolumn{3}{c}{\it Calculated light output and related variables} \\
%	&	&	\\
$C_{e,n}$ & constants in the approximative expressions of $S_{e,n}$ & a.u. \\
$a_1$ & gain fit parameter in the friendly analytical expression of L & a.u. \\
$a_2$ & quenching fit parameter in the friendly analytical expression of L &
a.u. \\
%$a_3$ & \shortstack{energy per nucleon threshold required to produce\\
%$\delta$ -- rays, a fit parameter in the friendly analytical expression of L} & MeV/$u$ \\
$a_3$ & $e_{\delta}$ fit parameter in the friendly analytical expression of L &
MeV \\
$a_4$ & fractional energy loss transferred to a $\delta$ - ray, a fit
parameter in the friendly analytical expression of L & \\
%$a_4$ & ${\cal F}$ fit parameter in the friendly analytical expression of L & \\
$f_{geom}$ & light collection factor & a.u. \\
$f_{PMT}$ & PMT gain factor & a.u. \\
%	&	&	\\
%\multicolumn{3}{c}{\it Other variables} \\
%	&	&	\\
%$h\nu$ & scintillation energy & eV \\
%$\cal T$ & absolute temperature & $^{\circ}$K \\
%$\Delta W$ & activation energy & eV \\
\hline
\end{supertabular}\\
%\end{tabular}\\
$^*$Most of the notations of the original references have been kept.
}
%%%%%%%%%%%%%%%%%%%%%%%%%%%%%%%%%%%%%%%%%%%%%%%%%%%%%%%%%%%%%%%%%%%%%%%%%%%%
%%%%%%%%%%%%%%%%%%%%%%%%%%%%%%%%%%%%%%%%%%%%%%%%%%%%%%%%%%%%%%%%%%%%%%%%%%%%
\section{The thallium-activated caesium iodide scintillators\label{sect2}}
%%%%%%%%%%%%%%%%%%%%%%%%%%%%%%%%%%%%%%%%%%%%%%%%%%%%%%%%%%%%%%%%%%%%%%%%%%%%
\subsection{CsI(Tl) crystals of INDRA and associated 
electronics\label{subsect21}}
%The 336 independent cells of INDRA are placed on 17 rings 
%(2$^\circ$ $<$ $\theta$ $<$ 176$^\circ$) \cite{pou95}. Forward cells 
%(rings:  2 -- 9,  covering 3$^\circ$ $<$ $\theta$ $<$ 45$^\circ$) have a
%"sandwich" structure of  three layers: gas ionisation chamber (IC),
%300 $\mu$m thick silicon detector (Si)  and thallium-activated
%caesium iodide scintillator (CsI(Tl)). Backward cells (rings 10 -- 17,
%covering 45$^\circ$ $<$ $\theta$ $<$ 176$^\circ$) have two layers
%only: IC and CsI(Tl). For energy calibration purpose, rings 10 -- 17
%were each equipped with a single two-element telescope (80 $\mu$m and
%2 mm thick silicon detectors) - which will be referred to as the
%calibration telescope (CT). The CT covers part of one of the CsI(Tl)
%crystals in each backward ring.
\par
There are 324 CsI(Tl) crystals in all, with thicknesses ranging between 138 mm
and 50 mm from forward to backward angles \cite{pou95}. All the 
crystals belonging to one ring have identical shape and size. Because the light
output of the CsI(Tl) crystals critically depends on the temperature
\cite{wil66,str90}, a water cooling system stabilizes at 20$^\circ$C the
temperature in the mechanical supports of the CsI(Tl).
%\par
The CsI(Tl) crystals are coupled to photomultipliers tubes (PMT)
\cite{pou95}. The use of PMTs provides
lower energy thresholds for mass identification as compared to those
obtained with photodiodes \cite{gui89}.
%\par
The stability control of the scintillators is ensured by optically
connecting them to a laser system \cite{pou95}, which makes use of a nitrogen
laser \cite{laser}, emitting in the ultraviolet (UV) band, and CsI(Tl) light
distributors.  
%\par
The CsI(Tl) PMT signals are fed in 24 input VXI bus modules containing the 
processing functions. Each channel comprises a constant fraction discriminator,
two integrators for ``fast'' and ``slow'' %components 
parts with accompanying delay
and gate generators. The analog to digital conversion is performed by
two multiplexed 12 bit converters.
%\par
Exhaustive descriptions of the CsI(Tl), PMT and associated electronics, as 
well as of the data acquisition and triggering system
are given in \cite{pou95,pou96}. We are doing here only a short presentation, 
stressing those details which are necessary to make comprehensible
the procedures concerning the CsI(Tl) energy calibration and the
fragment identification in the IC -- CsI(Tl) maps of the backward
rings. For energy calibration purpose, rings 10 -- 17
were each equipped with a single two-element telescope (80 $\mu$m and
2 mm thick silicon detectors) - which will be referred to as the
calibration telescope (CT). The CT covers part of one of the CsI(Tl)
crystals in each backward ring.
%%%%%%%%%%%%%%%%%%%%%%%%%%%%%%%%%%%%%%%%%%%%%%%%%%%%%%%%%%%%%%%%%%%%%%%%%%
\subsection{Calculation of the deposited energy into the
scintillators\label{subsect22}}
For rings 2 - 9, the calculation of the energy deposited into the CsI(Tl)
crystal to which a given light output corresponds - $E_{\mathrm 0}$ - is 
based on the energy lost in the preceding layer of a detection cell of INDRA, 
using the nominal thickness of each silicon detector ($\approx300 \mu$m).
The silicon detectors were carefully calibrated taking into account
the pulse height defect, with an absolute accuracy of $2 - 3\%$.
\cite{tab96}. However, the relative accuracy
- between different ions up to Xe and for different energies up to 80 AMeV -
is within $1\%$. For ions as light as Boron, the consequent relative accuracy 
for the residual energy in CsI(Tl) crystal is $1.3\%$ at 50 AMeV, $1.4\%$ at 
25 AMeV and $3\%$ at 5 AMeV.
For ions as heavy as Xenon, the consequent relative accuracy for the residual
energy in CsI(Tl) crystal is $1.8\%$ at 50 AMeV, $2.5\%$ at 25 AMeV and $10\%$
at 5 AMeV. One has to note that the relative accuracy of the total energy of
the ion, deposited in both Si and CsI(Tl) detectors never exceeds $2 - 3\%$.   
Starting from the energy deposited in the $300 \mu$m Si detector, and by
using the stopping power tables of Hubert et al. \cite{hub90} above 2.5 AMeV
and the renormalized variant of those of Northcliffe and Schilling \cite{nor} 
under 2.5 AMeV, the residual energy deposited in the scintillator placed 
behind it is calculated. For light charged particles and light fragments 
($Z\leq 4$), the mass is identified, while for heavier fragments, an 
hypothesis is necessary for the mass.
\par
For rings 10 - 17, %calibration telescopes (CTs) \cite{pou95} 
the CTs play the essential 
role in estimation of the energy deposited into the scintillators, as providing
reference energy spectra.  Instead of directly looking for the correspondence: 
total light output - deposited energy into the scintillator, event by event, 
the incremented spectrum (for each reaction product) is compared to the 
associated reference one, to which it is stretched. 
The calibration procedure for the backward rings is presented in subsection
\ref{subsect44}, followed by a description of a rapid fragment identification 
recipe making use of the same mentioned expression.
\par
All along this paper, the Xe + Sn system at 32 and 50 AMeV incident energies
is used and emitting reaction
products over a large range, in energy and atomic number, have been selected.
The experimental data were taped only for multiplicities %M 
higher than or equal to 4 %($M \geq 4$). 
(multiplicity $\geq 4$). Thus, the reaction products in the present
application originate only in exit channels involving a non negligible 
transfer of kinetic energy into internal degrees of freedom. As a consequence, 
the emitted primary fragments have excitation energies (greater than around 
1.5 AMeV) leading mainly to neutron evaporation. The secondary fragments 
(after evaporation) populate the ``attractor'' line in the map of nuclides 
\cite{cha98} rather than the ``stability'' line. The corresponding mass 
formula \cite{cha98} will be consequently employed when the isotopic
mass was not determined.
%Only 
In most of the cases, one CsI(Tl) detector will be used (module 2 of ring 3) 
to illustrate the described procedure. The nominal thickness of the preceding 
silicon detector is $304 \mu$m. 
%%%%%%%%%%%%%%%%%%%%%%%%%%%%%%%%%%%%%%%%%%%%%%%%%%%%%%%%%%%%%%%%%%%%%%%%%%
\subsection{The shape of the signal\label{subsect23}}
The light emitted by a CsI(Tl) crystal hit by a charged reaction product has a 
rise time negligible \cite{lia74,gra85,val93} as compared to the decay time
which is in the microsecond range. The rise time is related to the 
transfer of the energy deposited by the particle to the optical level involved 
into the scintillation, while the decay time constant concerns the light emission. 
\par
Traditionally, the decaying part of a CsI(Tl) scintillation is described by
one \cite{rob61ii,rob61i} or two \cite{gra85} exponentials of short 
decay-constant ($\approx 1\mu$s) or, more often, by one short decay-constant 
exponential ($\approx 1\mu$s) and one long decay-constant exponential 
(7 $\mu$s \cite{sto58}). The short decay-constant depends on the identity of
the particle, while the long one is considered to be the same for all 
particles.
\par
 Proton induced signals recorded by means of flash ADCs up to 20
 $\mu$s \cite{ben89} have shown decaying parts which are curved in a
 semilogarithmic scale, with a steep descent in the first 3-4 $\mu$s,
 which includes the larger part of the integral of the signal. The lower
 is the incident energy, the steeper is this part of the signal and the higher
 is its weight in the signal integral. For very low energy %($\approx$ 1 MeV),
 (a few MeV), the shape of %this first part of 
 the decaying signal is a straight line in
a semilogarithmic representation versus time, i.e. it shows an exponential
 decaying time dependence. At higher energy, several exponentials would be necessary for a perfect
 description of the shape of the whole decaying curve. As in Ref. \cite{sto58},
 the authors of Ref.~\cite{ben89} have kept only two. The first one -- the 
 ``fast component" -- for the dominant steep descent part, has a short decay-constant 
 (0.5 $\mu$s -- 1 $\mu$s), with a strong dependence on the atomic
 number Z, mass number A and incident energy $E_0$ of the particle. The second 
 one -- the ``slow component" -- has a long decay-constant
 (5 $\mu$s \cite{ben89}) nearly independent of the type of particle.
 \par
\begin{figure}[t]
\epsfxsize=10.cm
\epsfysize=10.cm
%\epsfbox{semnalkr.eps}
\epsfbox{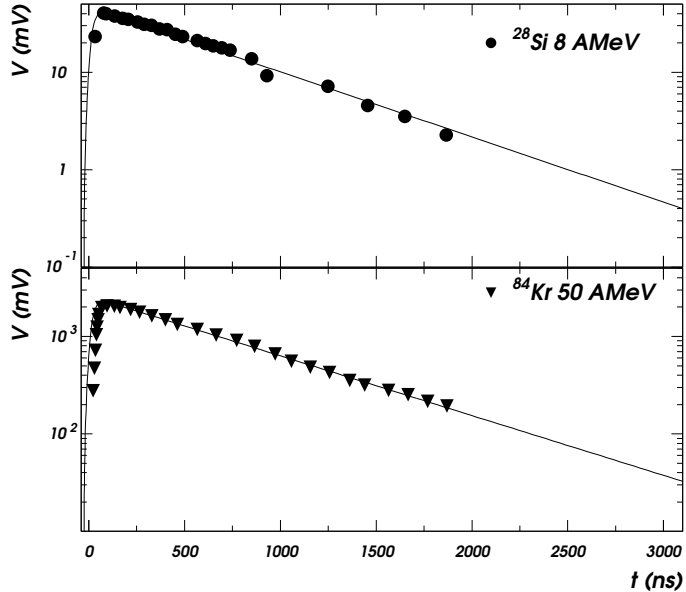}
\caption{The shape of the measured light signals induced by 8 AMeV $^{28}Si$  
        and 50 AMeV $^{84}Kr$ ions in two different CsI(Tl) crystals of INDRA 
	- solid symbols - is well described by one exponential decaying curve
	provided by Eq. (\ref{ecu3}).}
\label{fig1}
\end{figure}

 Pulse shape analysis allows particle discrimination up to Z=5
 for the  CsI(Tl) scintillators of INDRA. At the highest energies in Ref.
 \cite{ben89}, the fast component covers at least $\approx$ 65 \% of the
 integral of the signal for hydrogen isotopes at 20 - 40 AMeV,
 $\approx$ 75 \% for helium ones at 30 AMeV, $\approx$ 85\%
 for light fragments ($Z$=3-6) at 15 - 25 AMeV \cite{ben89}
 and at least 95 \% for heavy fragments, as shown in Fig. \ref{fig1} for
 Si at 8 AMeV and Kr at 50 AMeV (present work). The above mentioned weights
 become even more important when the incident energy $E_0$ decreases, i.e.
 when the average specific electronic stopping power $E_0/R(E_0)$ (for which a good 
 estimate, within a mutiplicative factor, is $AZ^2/E_0$) is high enough. 
 $R(E_0)$ is the corresponding particle range, and the estimate of the 
 average specific electronic stopping power is derived from the approximation of 
 Bethe-Bloch formula  $-(dE/dx)_{\mathrm e}\propto AZ^2/E$.  
%which leads to $E_0/R(E_0)\propto 2AZ^2/E_0$. 
 From the above considerations, one may assume
 in the latter case that only one decay-constant $\tau_0$($E_0$,A,Z)
 is involved for each event. Its value will be close to that of the fast 
 component.
%%%%%%%%%%%%%%%%%%%%%%%Figura 1%%%%%
%%%%%%%%%%%%%%%%%%%%%%%%%%%%%%%%%%%%
\par
 A signal at the crystal PMT output, described as in Ref. \cite{ben89} by two 
 exponential functions associated to the fast and slow component respectively:
 \begin{equation}
 i(t)=\frac{Q_{\mathrm f}}{\tau_{\mathrm f}} \e^{-\frac{t}{\tau_{\mathrm
 f}}} +
 \frac{Q_{\mathrm s}}{\tau_{\mathrm s}} \e^{-\frac{t}{\tau_{\mathrm s}}} ,
\label{ecu1}
\end{equation}
 has the total light response corresponding charge
 $Q_{\mathrm fs}=Q_{\mathrm f} + Q_{\mathrm s}$ got by
 integrating $i(t)$ over time between 0 and $\infty$.
 \par 
 In one exponential approximation, the same signal would be:
 \begin{equation}
 i(t)=\frac{Q_0}{\tau_0} \e^{-\frac{t}{\tau_0}} ,
\label{ecu2}
\end{equation}
 where $Q_0$ approximates the integral of the signal, $Q_{\mathrm fs}$.
%%%%%%%%%%%%%%%%%%%%%%%%%%%%%%%%%%%%%%%%%%%%%%%%%%%%%%%%%%%%%%%%%%%%%%%%%%
\section{Reconstruction of the total light output\label{sect3}}
\begin{figure}
\epsfxsize=12.cm
\epsfysize=12.cm 
%\epsfbox{test_1exp.eps}
\epsfbox{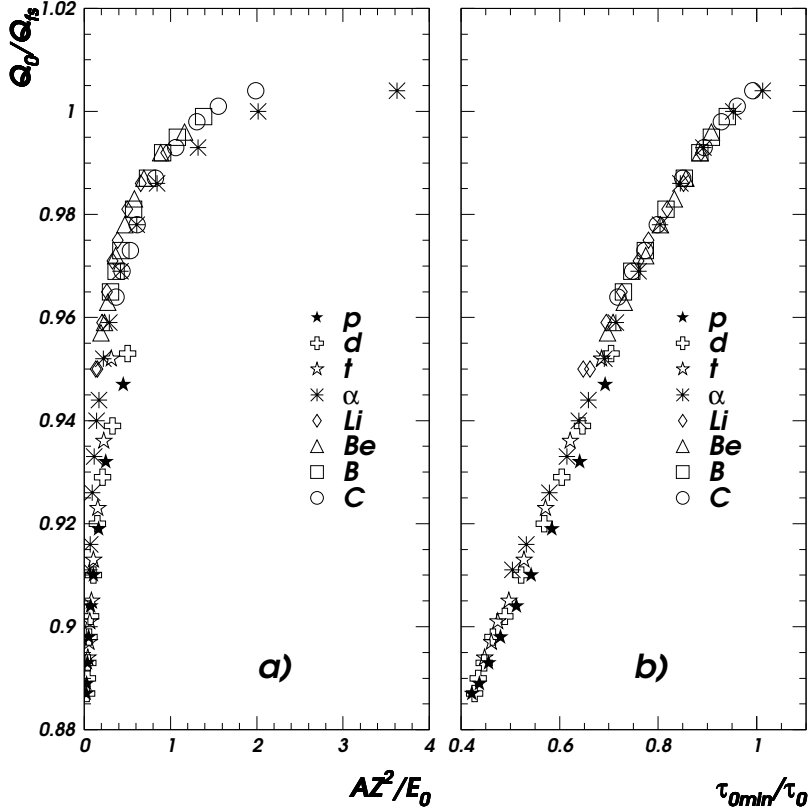}
\caption{$Q_0$ is the approximate integral of the light signal when one
	decaying exponential shape, of decay-constant $\tau_0$
	(Eq. \ref{ecu2}), is assumed. $Q_{fs}$ is the true one, corresponding
	to two decaying exponential shape (Eq. (\ref{ecu1})). Their ratio
	increases with: a) the estimate of the average specific electronic stopping
	power $\propto AZ^2/E_0$ ($E_0$ in MeV), b) the inverse ratio of 
	$\tau_0$; $\tau_{\mathrm 0min}$ is a normalization constant.
	%, given by the lowest value of $\tau_0$. 
	Processed data from Ref. \cite{ben89}.}
\label{fig2}
\end{figure}

 In the case of INDRA, only parts $F$, $S$ of the signal are integrated in the time
 gates 0 -- 400 ns and 1600 -- 3100 ns, respectively. Let us make the
 following exercise: consider the expressions of $F$ and $S$ provided by the
 two exponential formula of the signal (\ref{ecu1}) on one hand, and
 by the one exponential formula (\ref{ecu2}) of the signal, on the other hand.
 The values of the expressions of F and S found in both cases have to be 
 equal. From these equalities,  one can derive $\tau_0$ and
 the ratio $Q_{\mathrm 0}/Q_{\mathrm fs}$ for the data in Ref. \cite{ben89},
 but making use of the time gates of INDRA. The results of this estimation are
 plotted in Fig \ref{fig2}a) against the estimate $AZ^2/E_0$ of the average
 specific electronic stopping power. This plot has a predictive character:
 the maximum error done in the integral of the signal estimation
 would be of about $\approx$ 10\% in the case of the most energetic protons
 but much lower for the charged reaction products with $Z > 1$.
 Fig. \ref{fig2}b) shows the ratio $Q_{\mathrm 0}/Q_{\mathrm fs}$ versus the
 reciprocal of the decay-constant value $\tau_0$ of the ion in question; the
 normalization constant $\tau_{0min}$ is in fact the lower measured value
 for ions at the lowest energies ($\approx$ 1 AMeV) and hence the 
 highest stopping powers. This kind of plot could eventually
 be used in order to correct $Q_0$. In any case, as long as $AZ^2/E_0$ $\geq$
 0.4 (e.g. Ar ions with $E_0/A$ $\leq$ 810 MeV/nucleon), $Q_0$ estimates
 $Q_{\mathrm fs}$ within 2\%. This is the case for most of our data.
%%%%%%%%%%5 Figura 2
%%%%%%%%%%%%%%%%%%%%%%%%%%%%
 \par
 In view of the above argument, we shall suppose in the following that 
 the current at the last dynode of the PMT associated to a CsI(Tl) 
 scintillator varies exponentially in time as in Eq. (\ref{ecu2}), 
where $Q_0$ is the total charge. This current, injected in the anodic
circuit %RC (with the decay-constant of the circuit $\tau$=RC) 
of the PMT, leads to the measured current at the output of the PMT base
$i_{\mathrm mes}(t)$, 
%=dq(t)/dt$, where $dq(t)$ is the infinitesimal charge at the output, 
 which may be expressed
%\cite{kno89,riv96}
\cite{kno89} by means of the equation:
\begin{equation}
%\frac{dq(t)}{dt}
i_{\mathrm mes}(t)=\frac{Q_0}{\tau_0-\tau}\left(\e^{-\frac{t}{\tau_0}} -
\e^{-\frac{t}{\tau}}\right) .
\label{ecu3}
\end{equation}
The total light output %$L$ 
is proportional to the total charge $Q_0$; $\tau$ 
and $\tau_0$ are the rise time and decay time constants respectively. 
$\tau$ has been measured
%determined from the electronic components R (the resistor) and C
%(the capacity) inside the base of the PMT 
for the bases of all PMT (60 ns for rings 11 - 16 and 20 ns for other 
rings). The shape of the signal given by equation (\ref{ecu3}) is shown in 
Fig. \ref{fig1}. By integrating expression (\ref{ecu3}) within the gates 
mentioned above, one obtains:
\begin{eqnarray}
\label{ecu4}
F &=& \frac{Q_0}{\tau_0-\tau}\left[\tau_0\left(1-\e^{-t_1/\tau_0}\right) -
\tau\left(1-\e^{-t_1/\tau}\right)\right] \\
S &\approx&
\frac{Q_0\tau_0}{\tau_0-\tau}\left[\e^{-t_2/\tau_0}-\e^{-t_3/\tau_0}\right] ,
\label{ecu5}
\end{eqnarray}
with $t_1$=390 ns, $t_2$=1590 ns and $t_3$=3090 ns (the integration
gates have undergone a diminution of 10 ns according to the delay of
the signal in the VXI cards). The reason why it was necessary to approximate
the decaying shape of the signal by a single exponential function of time -
eq. (\ref{ecu3}) - is that in INDRA experiments, only $F$ and $S$ (the
channels of the charge convertors corresponding to the two gates) are measured
and not the integral of the signal. Under this assumption and by means of the
two eqs. (\ref{ecu4}) and (\ref{ecu5}), it is possible to find the two unknown
quantities $\tau_0$ and $Q_0$ and therefore, within a multiplicative constant,
the total experimental light output. The measured resolutions (FWHM) of the 
CsI(Tl) crystals for Si of 7.86 AMeV and $\alpha$ particles of 21 AMeV are of
$\approx 3\%$ for F and $\approx 4\%$ for S. This leads to an accuracy of
$\approx 1.3\%$ for $Q_0$ above 10 AMeV. Below a total energy of 10 MeV the
accuracy progressively goes down to 3 - 4$\%$, which roughly corresponds to 
the measured resolution on the total light for $\alpha$ particles of 5 MeV. For
 protons of 21 MeV, the resolutions (FWHM) are: $\approx 4\%$ for F and 
$\approx 7\%$ for S, leading to an accuracy of $\approx 2\%$ for $Q_0$ above 
10 MeV. Below this energy, the accuracy progressively goes down to 6 - 7$\%$.  
%%%%%%%%%%%%%%%%%%%%%%%%%%%%%%%%%%%%%%%%%%%%%%%%%%%%%%%%%%%%%%%%%%%%%%%%%%
%\section{Applications\label{sect4}}
\section{Approximate formula from RNQM model\label{sect4}}
%%%%%%%%%%%%%%%%%%%%%%%%%%%%%%%%%%%%%%%%%%%%%%%%%%%%%%%%%%%%%%%%%%%%%%%%%%
\subsection{Analytical integration\label{subsect41}}
In practical situations, %the numerical integration in the fit procedure is
%prohibited as being long computing time consuming and 
an analytical integration of the total light output issued from the 
RNQM \cite{par00i} would be more suited. This is possible starting from 
the first order approximation for total light output formula (expression (18)
%(\ref{ecu26}) 
in the previous paper \cite{par00i}): 

\begin{eqnarray}
\label{ecu26}
\nonumber
L & = & a_{\mathrm G}\left[\int^{E_\delta}_0\frac{1}{1+
a_{\mathrm n}S_{\mathrm n}(E) + a_{\mathrm R}S_{\mathrm e}(E)}\times\frac{\d E}
{1+S_{\mathrm n}(E)/S_{\mathrm e}(E)}\right. \\
\nonumber
& + & 
\int_{E_\delta}^{E_0}\frac{1-{\cal F}(E)}{1+
a_{\mathrm n}S_{\mathrm n}(E) + a_{\mathrm R}S_{\mathrm e}(E)}
\times\frac{\d E}{1+S_{\mathrm n}(E)/S_{\mathrm e}(E)} \\
%\nonumber
& + &
\left.\int_{E_\delta}^{E_0}\frac{{\cal F}(E)\d E}{1+S_{\mathrm
n}(E)/S_{\mathrm e}(E)}\right] ,
\end{eqnarray}

if suitable approximations are made for the stopping powers,
%$-(dE/dx)_{\mathrm e}(E)$, $-(dE/dx)_{\mathrm n}(E)$, 
the concentration $N_{\mathrm n}(E)$ of the defects created by the incident 
fragment and the fractional energy loss ${\cal F}(E)$ deposited outside the 
primary column by the generated $\delta$ -- rays. All these quantities 
are discussed in the preceding paper \cite{par00i}. 
\begin{itemize}
\item [i)] For the specific electronic stopping power formula
of Bethe-Bloch, the usual approximation:
$(\d E/\d x)_{\mathrm e}(E) = C_{\mathrm e} AZ^2/E$,  reasonable above a few 
AMeV, may be used; here $C_{\mathrm e}$ is a constant including the
logarithmic term in the Bethe-Bloch formula, much more slowly varying than
1/E.
\item [ii)] The created defect concentration $N_{\mathrm n}(E)$ estimated by 
$N_{\mathrm Ruth}(E)$, is well approximated by neglecting the second term 
($\propto E^{-2}$) in equation %(\ref{ecu10}):
(4) of the preceding paper \cite{par00i}
$N_{\mathrm n}\propto AZ^2/E$.
\item [iii)] The specific nuclear stopping power, $(\d E/\d x)_{\mathrm n}(E)$, 
may also be roughly estimated by an $AZ^2/E$ behaviour, as shown in Fig. 
%\ref{fig4}b):
2b) of the preceding paper \cite{par00i}:
$\left(\d E/\d x\right)_{\mathrm n}(E) = C_{\mathrm n} AZ^2/E$ with
$C_{\mathrm n}$ constant. In this way, the factor
$(1+S_{\mathrm n}(E)/S_{\mathrm E}(E))$ in the denominator of all the terms in
expression (\ref{ecu26})  becomes a constant: $1+C_{\mathrm n}/C_{\mathrm e}$, 
to be included in the multiplicative parameter $a_{\mathrm G}$ that will be 
called $a_{\mathrm 1}$, and the nuclear and recombination quenching terms 
to the denominator of the first two terms (concerning the primary column) in 
the same expression %(\ref{ecu26}) 
%(18) of the preceding paper \cite{par00i} 
may be summed and replaced by only one: $a_{\mathrm 2}AZ^2/E$. 
%\item [ii)] The specific nuclear stopping power, $(\d E/\d x)_{\mathrm n}(E)$, 
%may also be roughly estimated by an $AZ^2/E$ behaviour, as shown in Fig. 
%%\ref{fig4}b):
%2b) of the preceding paper \cite{par00i}:
%$\left(\d E/\d x\right)_{\mathrm n}(E) = C_{\mathrm n} AZ^2/E$ with
%$C_{\mathrm n}$ constant. In this way, the factor
%$(1+S_{\mathrm n}(E)/S_{\mathrm E}(E))$ in the denominator of all the terms in
%expression (\ref{ecu26})  becomes a constant: 
%$1+C_{\mathrm n}/C_{\mathrm e}$, to be included in the multiplicative 
%parameter $a_{\mathrm G}$ that will be called $a_{\mathrm 1}$. 
%\item [iii)] The created defect concentration $N_{\mathrm n}(E)$ estimated by 
%$N_{\mathrm Ruth}(E)$, is well approximated by neglecting the second term 
%($\propto E^{-2}$) in equation %(\ref{ecu10}):
%(4) of the preceding paper \cite{par00i}
%$N_{\mathrm n}\propto AZ^2/E$. The nuclear and recombination quenching terms 
%to the denominator of the first two terms (concerning the primary column) in 
%expression (\ref{ecu26}) 
%%(18) of the preceding paper \cite{par00i} 
%may be thus summed and replaced by only one: $a_{\mathrm 2}AZ^2/E$. 
\item [iv)] By keeping the zero and first order terms in the Taylor expansion 
around $\beta_{\mathrm \delta}^2$ of the logarithmic term of the fractional 
energy carried by the $\delta$ -- rays (see Eq. %(\ref{ecu15}) 
(7) of the preceding paper \cite{par00i}), one may get an approximate 
expression of ${\cal F}(\beta^2)$:
\end{itemize}
\begin{equation}
{\cal F}(\beta^2)=\frac{1}{2}\frac{\frac{\beta^2}{\beta_{\mathrm \delta}^2}-1}
{\ln(\frac{2m_{\mathrm e}c^2}{I}\beta_{\mathrm \delta}^2)+
\frac{\beta^2}{\beta_{\mathrm \delta}^2}-1} .
\label{ecu27}
\end{equation}
With the above items i)--iv) assumptions, the first order approximation
formula (\ref{ecu26}) of the total light output depends on three
parameters only and may be analytically integrated.
%The corresponding values of the fit parameters $a_{\mathrm i}$, i=1,3, (we made 
%the notation $a_{\mathrm 3}=e_{\mathrm \delta}$) obtained for the same 
%illustrative module 
%3\_2 
%and for the same set of data as before with the same condition 
%($AZ^2/E_{\mathrm 0} \geq 0.4$) are given in Table \ref{table3}a).
%% TABLE 3 %%%%%%%%%%%%                                         %%%%%%%%%%%%%%
%\begin{table}
%\caption{Fit parameters $a_{\mathrm 1}$, $a_{\mathrm 2}$, $a_{\mathrm 3}$,
%	$a_{\mathrm 4}$: gain, quenching, $e_{\mathrm \delta}$ and ${\cal F}$ 
%	for module 2 of ring 3. Analyticaly integrated expressions of the 
%	total light output have been used: a) items i) - iv) in subsection 
%	\ref{subsect41}, 
%	b) Eq. (\ref{ecu29}). The recommended values of 
%	$a_{\mathrm 2}$, $a_{\mathrm 3}$, $a_{\mathrm 4}$ to be used in 
%	Eq. (\ref{ecu29}) - column c) are suitable for all INDRA CsI(Tl) 
%	crystals.}
%\begin{center}
%\begin{tabular}{cccc}
%\hline
% & a) &	b) &  c) \\
%$a_1$ & 9.09 &	16.38 & variable \\
%$a_2$ & 9.34$\times 10^{-2}$ & 0.386 &	0.25 \\
%$a_3$ & 2.50 & 3.99 &  3.10 (1.00)$^{*}$ \\
%$a_4$ & - &  0.284 & 0.270 \\
%\hline
%\end{tabular} \\
%\end{center}
%$^*$ see the text for explanation
%\label{table3}
%\end{table}
%% TABLE 3 %%%%%%%%%%%%                                         %%%%%%%%%%%%%%
\begin{table}
\caption{Fit parameters $a_{\mathrm 1}$, $a_{\mathrm 2}$, $a_{\mathrm 3}$,
	$a_{\mathrm 4}$. %gain, quenching, $e_{\mathrm \delta}$ and ${\cal F}$.
	The errors on the parameters (one unit on the last digit) are only
	statistical. The analytically integrated expression (\ref{ecu29}) of
	the total light output has been used: a) values obtained by
	means of data available in the whole ion range, for a forward module
	($\theta = 4.5^\circ$) of INDRA; b) values averaged over 8 modules
	placed on the forward rings ($\theta \leq 45^\circ$) obtained only by 
	means of intermediate mass fragments and light charged particles.
	These recommended values of $a_{\mathrm 2}$, $a_{\mathrm 3}$, 
	$a_{\mathrm 4}$ to be used in Eq. (\ref{ecu29}) are suitable for all 
	INDRA CsI(Tl) crystals.}
\begin{center}
\begin{tabular}{lccc}
\hline
 & a) &	b) \\
%$a_1$ & 10.74 &	19.50 & variable \\
%$a_2$ & 0.171 & 0.705 &	0.4 - 0.5 \\
%$a_3$ & 4.35 & 3.80 &  3.10 (1.00)$^{*}$ \\
%$a_4$ & - &  0.261 & 0.270 \\
$a_1$~[a.u.] &	19.5 & variable \\
$a_2$~[a.u.] & 0.71 &	0.25 \\
$a_3$~[MeV$/u$] & 3.8 &  3.1 (1.0)$^{*}$ \\
$a_4$ &  0.26 & 0.27 \\
\hline
\end{tabular} \\
\end{center}
$^*$ see the text for explanation
\label{table3}
\end{table}
%%%%%%%%%%%%%%%%%%%%%%                                          %%%%%%%%%%%%%%
The quality of the fragment loci reproduction in a 
%$\Delta E_{Si} - Light_{CsI(Tl)}$
$\Delta E_{Si} - Q_0$ map will be shown in the next 
subsection.
%The quality of the fit is shown in Figs.
%\ref{fig9},\ref{fig10},\ref{fig11} - dashed lines.
%%%%%%%%%%%%%% Figura 9 %%%%%%%%%%%%%%%%%%%%%%%%%%%%%%%%%%%%
\begin{figure}
\epsfxsize=12.cm
\epsfysize=12.cm
%\begin{rotate}{56} 
%\epsfbox{/tmp_mnt/import/projet2/indra3/tabacaru/csi/hvse_aprox.eps}
\epsfbox{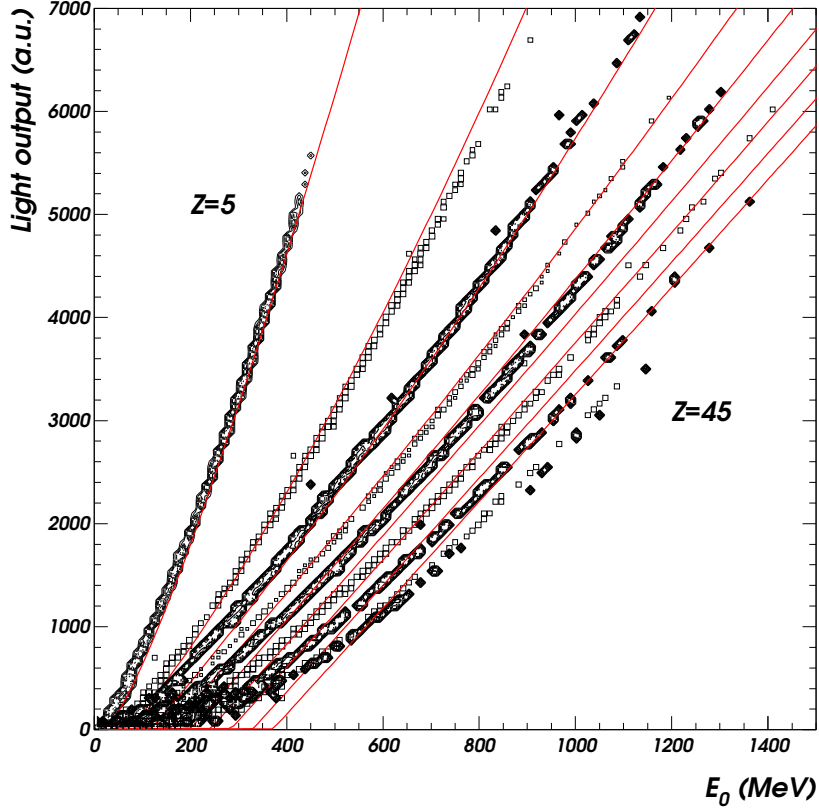}
%\epsfbox{hvse_aprox.eps}
%\end{rotate}
%\caption{The same as in Fig. \ref{fig6}, but when calculations are done with
%        the simple light output formula Eq. (\ref{ecu29}).}
\caption{Total light output against the energy for different ions 
	($\Delta Z$ = 5): the symbols are experimental data from the system 
	Xe + Sn at 32 and 50 AMeV; the curves are calculations done with the 
	simple light output formula Eq. (\ref{ecu29}).} 
\label{fig9}
\end{figure}
%%%%%%%%%%%%%% Figura 10 %%%%%%%%%%%%%%%%%%%%%%%%%%%%%%%%%%%%%%%%%
\begin{figure}
\epsfxsize=12.cm
\epsfysize=12.cm
%\begin{rotate}{56} 
%\epsfbox{/tmp_mnt/import/projet2/indra3/tabacaru/csi/hvsesi_aprox.eps}
\epsfbox{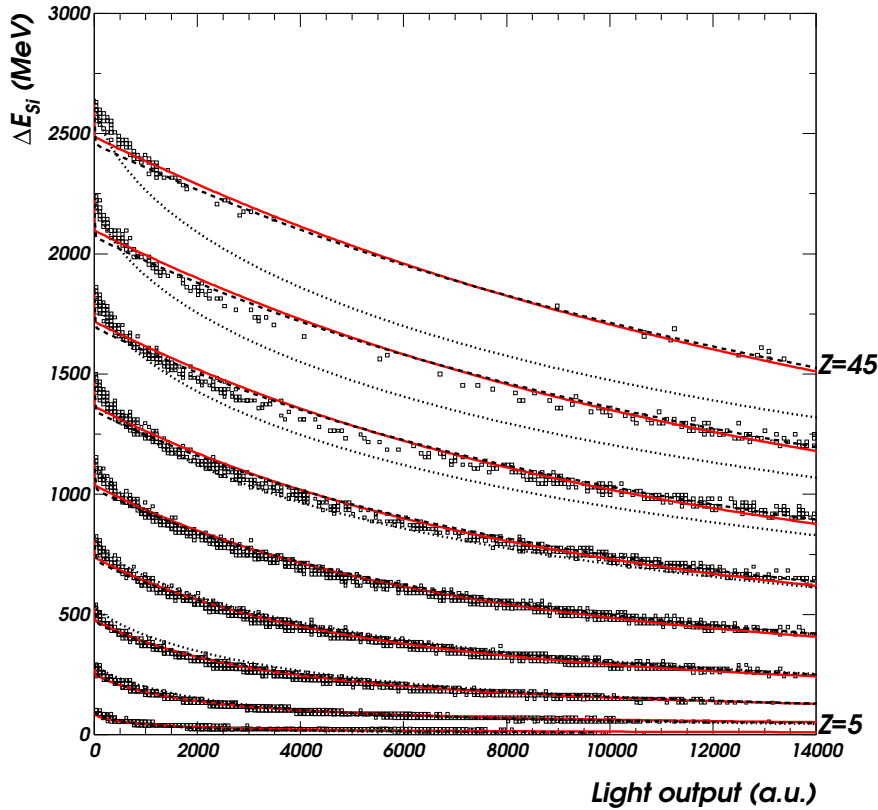}
%\epsfbox{hvsesi_aprox.eps}
%\end{rotate}
\caption{%The same as in Fig. \ref{fig7} as concerns the experimental data.
	A %$\Delta E_{Si} - Light_{CsI(Tl)}$
	$\Delta E_{Si} - Q_0$ map (module 2, ring 3 of
	INDRA) from the sytem Xe + Sn at 32 and 50 AMeV. The symbols are
	experimental data. For the calculated total light output, the 
	approximations i) - iv) from subsection \ref{subsect41} have allowed 
	an analytical integration, by making use of two approximate 
	expressions of ${\cal F}$: Eq. (\ref{ecu27}) - dashed lines - and the 
	step function Eq. (\ref{ecu28}) - solid lines; the dotted lines are 
	obtained if the $\delta$ -- rays are completely neglected 
	(${\cal F} = 0$) (subsection \ref{subsect45}).}
\label{fig10}
\end{figure}
%%%%%%%%%%%%%%%%%%%%%%%%%%%%
%%%%%%%%%%%%%% Figura 11
\begin{figure}
\epsfxsize=12.cm
\epsfysize=12.cm
%\begin{rotate}{56} 
%\epsfbox{/tmp_mnt/import/projet2/indra3/tabacaru/csi/ecartaprox.eps}
\epsfbox{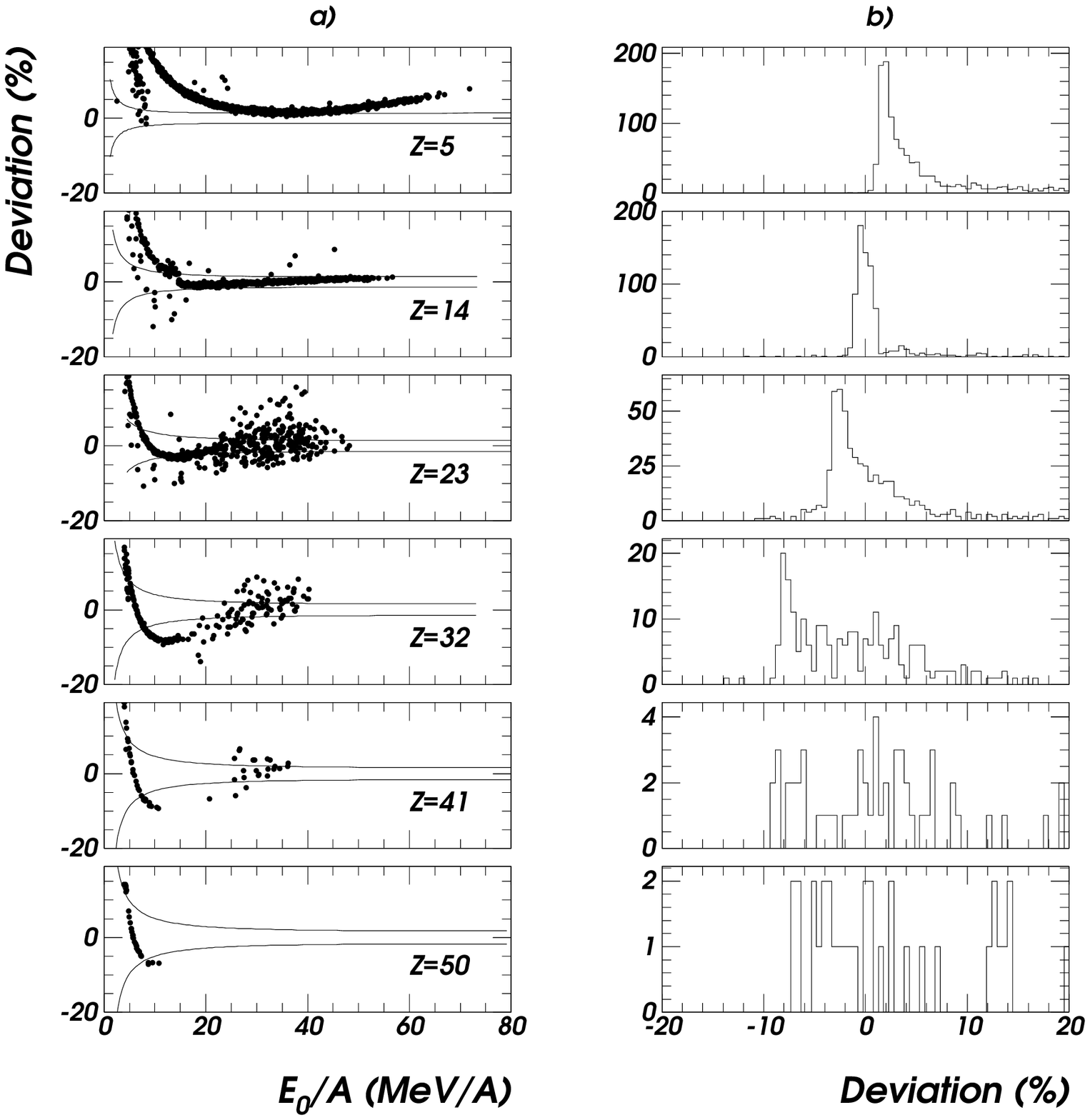}
%\epsfbox{ecartaprox.eps}
%\end{rotate}
%\caption{Same as Fig. \ref{fig8}, but when calculations are done with
%	the simple light output formula Eq. (\ref{ecu29}).}
\caption{Deviations (in \%) of the energy values, determined with the simple
	light output formula Eq. (\ref{ecu29}), from the true energy values
	for several products of the reactions Xe + Sn at 32 and 50 AMeV: a)
	deviations plotted against the product energy per nucleon: symbols;
	the regions between curves show the accuracy of the true energy per 
	nucleon (see subsection \ref{subsect22} for details); b) deviation
	histograms.} 
\label{fig11}
\end{figure}
%%%%%%%%%%%%%%%%%%%%%%%%%%%%%%%%%%%%%%%%%%%%%%%%%%%%%%%%%%%%%%%%%%%%%%%%%%%%%
\subsection{A friendly analytical formula for the total light
output\label{subsect42}}
The alternative to the item iv) approximation of the fractional energy
carried by the $\delta$ -- rays is to consider it as a step function of energy:
\pagebreak
\begin{equation}
{\cal F}(E)=\left\{\matrix{
0, & E/A \leq a_3 \cr
a_4, & E/A > a_3 \cr
}\right. ,
\label{ecu28}
\end{equation}
where $a_3$ is the energy per nucleon threshold for the $\delta$ --
ray production and $a_4$ will be a fit parameter too. The advantage is that 
the first order approximation of total light output expression (\ref{ecu26}) 
becomes a simple, easily to handle one:
\begin{equation}
L=a_1\left\{E_0\left[1-a_{\mathrm
2}\frac{AZ^2}{E_0}\ln\left(1+\frac{1}{a_2\frac{AZ^2}{E_{\mathrm
0}}}\right)\right]+a_4a_2 AZ^2
\ln\left(\frac{E_0+a_2AZ^2}{E_\delta+a_2AZ^2}
\right)\right\}
\label{ecu29}
\end{equation}
($E_{\delta}=A\times a_3$), very suitable for energy calibration
purposes. The fit parameter values are given in column a) of Table 
\ref{table3}, and the quality of the fit is shown in Figs. 
\ref{fig9},\ref{fig10} (solid lines) and \ref{fig11}. Even if the total light
outputs are no more as nicely reproduced as by exact calculations
\cite{par00i}, especially for %very 
high specific electronic stopping power values, the description of the 
reaction product identification in the two-dimensional plot (Fig. \ref{fig10}) 
and the deviations of the calculated energies relative to the true energies 
(Fig. \ref{fig11}) remain comparable to the exact calculation case.
More precisely, the heavy fragment identification in a
%$\Delta E - Light_{CsI(Tl)}$
$\Delta E - Q_0$ map is possible with a resolution of one 
unit charge around $Z$=40. The corresponding energy deviations may
locally reach up to  15\% - 20\%, but globally there are inside
$\approx 6\%$. About 3\% of accuracy are lost as compared to the exact 
calculations.
Note in Fig. \ref{fig10} that the step function approximation for 
$\cal F$ (solid lines) does not worsen the result as compared to the physical 
approximation (\ref{ecu27}) (dashed lines). 
In fact, it is approximation i) for the Bethe-Bloch 
formula which is distorting - via the fit procedure - the shape of the light 
response for the heaviest fragments (Fig. \ref{fig9}).
%%%%%%%%%%%%%%%%%%%%%%%%%%%%%%%%%%%%%%%%%%%%%%%%%%%%%%%%%%%%%%%%%%%%%%%%%%%%
\par
For the energetic light charged particles, there are discrepancies at
high energy, whose origin is the slight underestimation of total light output
(see sect~\ref{subsect23}). For energy calibration purpose, the recipe to
ameliorate the situation was to use a different gain parameter $a_1$ %$a_{\mathrm G}$ 
for low average specific electronic stopping power ($(AZ^2/E_{\mathrm 0}<0.4$), 
or directly for protons, which constitute most of the data for which the 
experimentally determined $Q_{\mathrm 0}$ differs from the real experimental 
light output by $\approx 2 - 10\%$ (Fig. \ref{fig2}a)). The alternative could 
be a previous correction of $Q_{\mathrm 0}$ concerning these data, based on the 
$\tau_{\mathrm 0min}/\tau_{\mathrm 0}$ values, as suggested by 
Fig. \ref{fig2}b).
\par
\begin{figure}
%\epsfxsize=12.cm
%\epsfysize=12.cm
%\epsfbox{/tmp_mnt/import/projet2/indra3/tabacaru/csi/param_zmax_stabil.eps}
\epsfbox{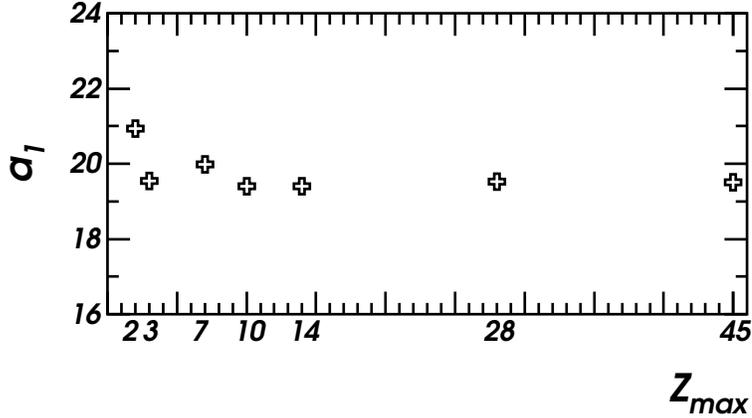}
%\epsfbox{param_zmax_stabil.eps}
\caption{Variation of the parameter $a_1$ as a function of the upper limit
of Z considered for fit; $a_2$, $a_3$, $a_4$ were kept fixed with values
from Table \ref{table3}a). The error bars are smaller than the symbols because only
statistical errors on the parameters have been considered.}
\label{fig_new}
\end{figure}

An interesting point is that the gain parameter values got by 
using only light fragments, or even light charged particles are very near the 
values obtained by using a large ion data range up to $Z$=45, if the 
other parameters, provided by the whole range of ions, are kept constant at the 
 values given in Table \ref{table3}a). Such a conclusion clearly appears in 
 Fig. \ref{fig_new}: $a_1$ values are found nearly independent of the upper 
limit in Z ($Z_{max}$) considered, when the last three parameters are fixed . 
If only $a_3, a_4$ are kept constant at the mentioned values,
$a_1, a_2$ remain practically the same as long as $Z_{max} \geq$ 14. Otherwise,
as the light response induced by intermediate mass fragments or light charged 
particles is less non-linear than that corresponding to heavy ions, the parameter $a_2$ has the 
tendency to diminish and, consequently, $a_1$ too, for the same quality of the 
description. 
%%%%%%%%%%%%%% Figura new
%%%%%%%%%%%%%%%%%%%%%%%%%%%%
\par
A search of the fit parameter values was performed for one
module of each of the 8 forward rings equipped with preceding
silicon detectors, by using as input data the light outputs induced by
intermediate mass fragments, accessible in the covered angular
domain ($\theta \leq 45^\circ$) for the studied system. It has led to very
similar values of the parameters $a_3, a_4$, close to those provided by large
$Z$ and $E_0$ scale data. The quenching parameter $a_2$ is related
to the average stopping power of the reaction product, i.e. to its identity 
and energy, but also to the activator concentration of the crystal, as we
shall see in the next subsection. It may vary from one to another ring, but
not dramatically. The corresponding averaged values of these 3 parameters - 
presented in column b) of Table \ref{table3} - are the recommended values for 
the applications when formula (\ref{ecu29}) is used. $a_3$ was decreased at 
the lower detection threshold (the value in parenthesis) in order to avoid the 
discontinuity in the energy spectra induced by the ``non-derivability'' at 
$E = E_{\mathrm \delta}$ of the light output expression as a function of the 
energy. By keeping $a_2, a_3, a_4$ fixed, the remaining gain parameter may be 
accurately determined as a free parameter by using simply light charged 
particles. The results are very similar to those shown in Figs. 
\ref{fig9},\ref{fig10} with solid lines and in Fig. \ref{fig11}. 
%with solid symbols. 
%%%%%%%%%%%%%%%%%%%%%%%%%%%%%%%%%%%%%%%%%%%%%%%%%%%%%%%%%%%%%%%%%%%%%%%%%%
%%%%%%%%%%%%%%%%%%%%%%%%%%%%%%%%%%%%%%%%%%%%%%%%%%%%%%%%%%%%%%%%%%%%%%%%%%%%%%
\subsection{Comparative study of the CsI(Tl) crystals of
INDRA\label{subsect43}}
\begin{figure}
\epsfxsize=12.cm
\epsfysize=12.cm
%\begin{rotate}{56} 
%\epsfbox{/tmp_mnt/import/projet2/indra3/tabacaru/csi/a1a2gain.eps}
\epsfbox{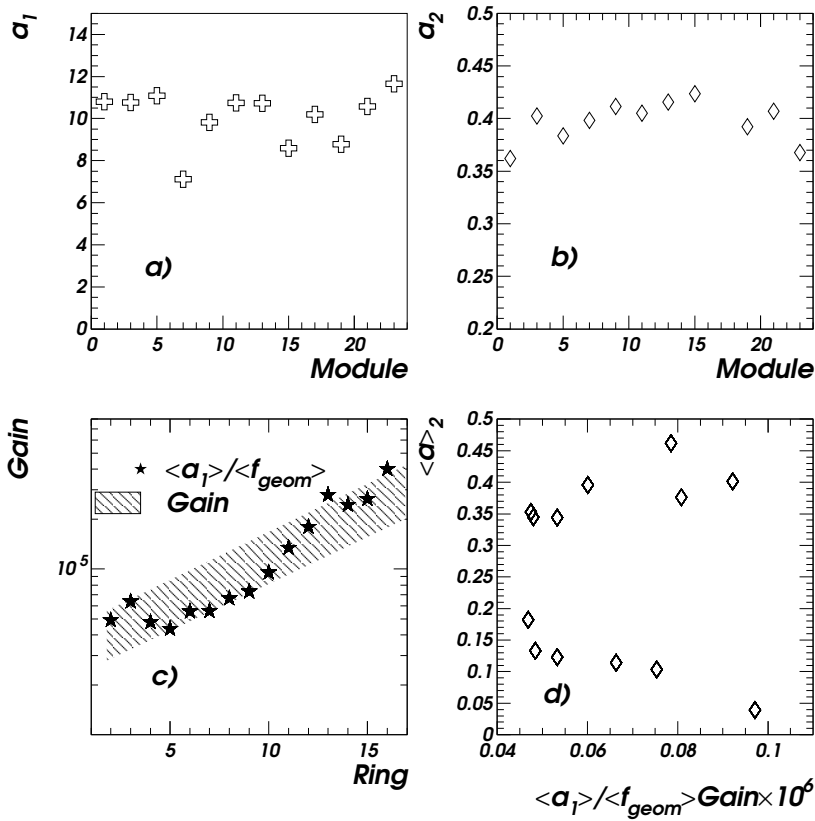}
%\epsfbox{a1a2gain.eps}
%\end{rotate}
\caption{a), b) The fit parameters $a_1$, $a_2$ used in the simple formula
	(\ref{ecu29}) against the module number for ring 2 of INDRA.
	c) The parameter $<a_1>$, averaged over all the modules of each ring
	and corrected for the light collection, versus the ring number;
	it follows the PMT gains - hatched area. d) The quenching parameter
	$<a_2>$, averaged over the modules of each ring, versus the parameter 
	$<a_1>$ corrected for light collection and PMT gain, averaged over
	the modules of the same ring. The error bars are smaller than the
	symbols because only statistical errors on the parameters have been 
	considered.} %In the cases c),d), only light charged particle data have been
	%used}
\label{fig13}
\end{figure}

The total light output friendly formula (\ref{ecu29}) was used to perform a
comparative study of the CsI(Tl) crystals of INDRA. 
% The fit procedure contains no constraint concerning the average specific 
%electronic stopping power.
At forward angles (rings 2 - 9), the experimental values were obtained from 
elastically scattered light %and heavy 
 ions on various targets (C, Al, Au). The light ions were produced by secondary
 beams with atomic number Z = 1 - 6 and mass number A precisely identified. 
The threshold energy per nucleon $e_{\mathrm \delta}$ required to generate 
$\delta$ -- rays which contribute to the scintillation is an intrinsic 
characteristic of the CsI crystals. Consequently, after a grid search
over all the modules of the 8 forward rings, the parameters 
$a_{\mathrm 3}$, and hence $a_{\mathrm 4}$ too, were fixed. The crystals 
belonging to one ring have the same size, the same shape and similar Tl (as 
well as eventually defect) concentrations. The gains of the associated PMT are 
the same. For this reason, one would expect that the parameters $a_1$ and 
$a_2$:
\begin{displaymath}
a_{\mathrm 1}\propto f_{\mathrm geom}\times f_{\mathrm PMT}\times 
\frac{\Laena}{\Laena+\Ldnd},~~~~~~~a_{\mathrm 2}
\propto\frac{1}{\Laena+\Ldnd},
\end{displaymath}
take nearly the same values for the modules of one ring; the quantities
$\Laena$, $\Ldnd$ were defined in the previous paper \cite{par00i} and
$f_{\mathrm geom}$, $f_{\mathrm PMT}$ are factors connected to the light 
collection (geometry of the crystal) and to the PMT gain, respectively.  
Actually, this is the situation, as shown in Fig. \ref{fig13}a),b) for the 
modules of ring 2.
%%%%%%%%%%%%%% Figura 13
%%%%%%%%%%%%%%%%%%%%%%%%%%%%
\par
If $\Ldnd/\Laena\ll 1$, the parameter $a_{\mathrm 1}/f_{\mathrm geom}$ is
mainly related to the associated PMT gain. The geometrical light collection
factor $f_{\mathrm geom}$ is proportional to the response of the crystal at
$^{137}Cs$ source $\gamma$ -- ray irradiation measured with the same PMT for
all the crystals of INDRA. Averaged over the modules of the same ring, the 
parameter $<a_1>/<f_{\mathrm geom}>$ plotted versus the ring number, follows 
the approximately known values of the PMT gain (provided by the manufacturer) 
as shown in Fig. \ref{fig13}c). The correlation of the two parameters 
$<a_1>$ and $<a_2>$ may be followed in Fig. \ref{fig13}d), if $<a_1>$ is 
previously corrected for the geometric and gain factors. Obviously, there is 
no mathematical correlation: two groups of detectors are put in evidence. They 
may correspond to the different concentrations of the Tl activator -  
from 200~ppm~ to 2000~ppm~ - (and eventually of the other crystal 
imperfection) which could appear during the CsI(Tl) crystal growth 
through the Bridgman method \cite{bdh}.
%In fact, the quantity $<a_1>/(<a_2><f_{\mathrm geom}><f_{\mathrm PMT}>)$ was 
%chosen as representation ordinate, because, after the above parameter formulae,
% it would be propotional to the Tl concentration. It seems that, during the 
%Csi(Tl) crystal growth \cite{ref ? Pouthas}, a weak Tl concentration
%gradient would be induced along the huge initial ingots. The crystal
%dedicated to a given ring of INDRA, were always cut at the same coordinate on 
%the ingot lenghts, to ensure quite similar Tl concentration. From ring to 
%ring, it could eventually exist slight activator concentration differences, 
%within the same order of magnitude. If this is true, the figure \ref{fig13}d) 
%would account not only for the mathematical correlation of the two parameters, 
%but it would have psysical meaning: the higher would be the activator 
%concentration (below the saturation domain mentioned in subsection 
%\ref{subsect31}, the lower would be the quenching parameters.  
%During the Csi(Tl) crystal growth \cite{ref ? Pouthas}, a Tl concentration gradient is induced 
%along the huge initial ingots. The crystal dedicated to a given ring of 
%INDRA, were always cut at the same coordinate on the ingot lengths, to ensure 
%quite similar Tl concentration. After the above parameter formulae, the 
%quantity $<a_1>/(<a_2><f_{\mathrm geom}><f_{\mathrm PMT}>)$ would be 
%propotional to the Tl concentration, and it is plotted versus the ring number 
%in Fig. \ref{fig13}d). Actually, it seems to indicate small activator 
%concentration variations, within the same order of magnitude. 
At backward angles of INDRA (rings 10 - 17), only light charged particles,
available from secondary beams at GANIL, could be used in the fit procedure to 
get the results presented in Fig. \ref{fig13}c),d). For unitarity, in these last
two plots, we have also restricted the forward angle employed data to light 
charged particles only. 
%At backward angles of INDRA (rings 10 - 17), only light charged particles, 
%available from secondary beams at GANIL, 
%were used in the fit procedure to get 
%the results presented in Fig. \ref{fig13}. 
\par
The above statistics performed over more than 300 CsI(Tl) crystals belonging 
to INDRA array and the related considerations enforce the consistancy of
the RNQM and of the exact or approximate light output expressions obtained in
its framework. %The CsI(Tl) response saturates very rapidly with the Tl
%concentration for the $\gamma$ - ray and light charged particle
%induced scintillations \cite{gwi63i,eby54}. The available crystals normally
%have activator concentrations in the saturation domain for these
%reaction products. Even if, for intermediate mass fragments and haevy ions,
%this request seems to be
%not fully accomplished, the quenching parameter $a_2$ in Eq. (\ref{ecu29}),
%or at least its order of magnitude, could be determined and an average
%value, reasonable for all the modules of INDRA, could be proposed (column b)
%of Table \ref{table3}), as shown in the previous subsection. Together with the
%other two: $a_3,a_4$, the intrinsic CsI parameters related to $\delta$ -- rays
%, it may characterize
The parameters $a_3,a_4$ and, to a certain extent, $a_2$ may be 
considered as intrinsic characteristics of the CsI(Tl) scintillators used in 
nuclear physics applications. Insofar, the rather simple formula (\ref{ecu29}) 
of the total light output, together with the recommended values in Table 
\ref{table3}b) constitute a powerful tool to be used in heavy ion experiments. 
Starting from easily to obtain light charged particle experimental data 
(excluding protons if their total light signal is not directly measured), they 
allow to accomplish useful applications such as heavy fragment identification 
and their energy determination, in case of lack of directly related 
information. An example is given in the following subsection.  
%%%%%%%%%%%%%%%%%%%%%%%%%%%%%%%%%%%%%%%%%%%%%%%%%%%%%%%%%%%%%%%%%%%%%%%%%%%%%
\subsection{Reaction product identification and energy calibration of the
backward angle modules of INDRA\label{subsect44}}
There are two major difficulties when the CsI(Tl) scintillators of INDRA placed
at polar angle above $45^\circ$ are exploited. Firstly, 
%at backward angle (rings 10 -- 17) the cells of INDRA have
%only two detection layers: IC and CsI(Tl) scintillators. 
in a $\Delta E_{\mathrm IC} - Q_0$ map the identification is perfect for low
atomic number $Z$ of the fragments ($Z\le 15$). For higher $Z$ values
the identification of the atomic number becomes uncertain, because of
the poor statistics and the lower IC energy resolution, as compared to the Si
detectors. Secondly, at large angles it is quite difficult to obtain data of
elastic scattering %in normal kinematics 
for fragments. Insofar, the calibration of the module belonging to the same 
ring is performed in two stages.
\par
In the first stage, we %deal with 
consider the CsI(Tl) partially obturated by the
calibration telescope CT (Si(80 $\mu$m) - Si(Li) 2 mm), one per ring.
%For the Si(Li) detectors which are not reached by the $\alpha$ particles
%of a thoron source, the absolute energy is obtained from secondary
%proton and $\alpha$ beams.
Most of the fragments ($Z\geq$4) are stopped in the Si(Li) and therefore
identified in the Si -- Si(Li) telescope.
The corresponding energy spectrum is built for
each atomic number $Z$. There are energetic light charged particles
($ > 20$ AMeV) passing through the
silicon telescope and stopped in the CsI(Tl) crystal coupled behind.
The calculated \cite{hub90,nor} residual energies are the ``true'' energies
deposited in the scintillator. Relied on the analytical expression
(\ref{ecu29}) of the light output, and using the recommended values from
Table \ref{table3}b) for parameters $a_2$,$a_3$ and $a_4$, the parameter 
$a_{\mathrm 1}$ is determined by a fit procedure. By means of these parameters,
 the energy spectrum in the scintillator is built for each of the light 
charged particles punching through the silicon telescope. Put together the 
silicon telescope and CsI(Tl) scintillator spectra provide the whole energy spectrum 
for a given particle. These spectra and those of fragments up to $Z$=15 
stopped in the Si -- Si(Li) telescope are the reference spectra for the 
respective ring, concerning an energy range: 
$E_{\mathrm min}, ~E_{\mathrm max}$ for each type of reaction product.
At the same time, the parameters  allow to perform $Z$ 
identification in a $\Delta E_{\mathrm IC} - Q_0$ map for the whole range of 
atomic number of light charged particles and fragments passing besides the 
silicon telescope and entering the CsI(Tl) coupled behind. 
%This fact is allowed by software constraints. 
The energy spectra of these reaction products were compared with the reference 
spectra; the superposition is quite good.
\par
\begin{figure}
\epsfxsize=12.cm
\epsfysize=12.cm
%\epsfbox{/tmp_mnt/import/projet2/indra3/tabacaru/csi/etire.eps}
\epsfbox{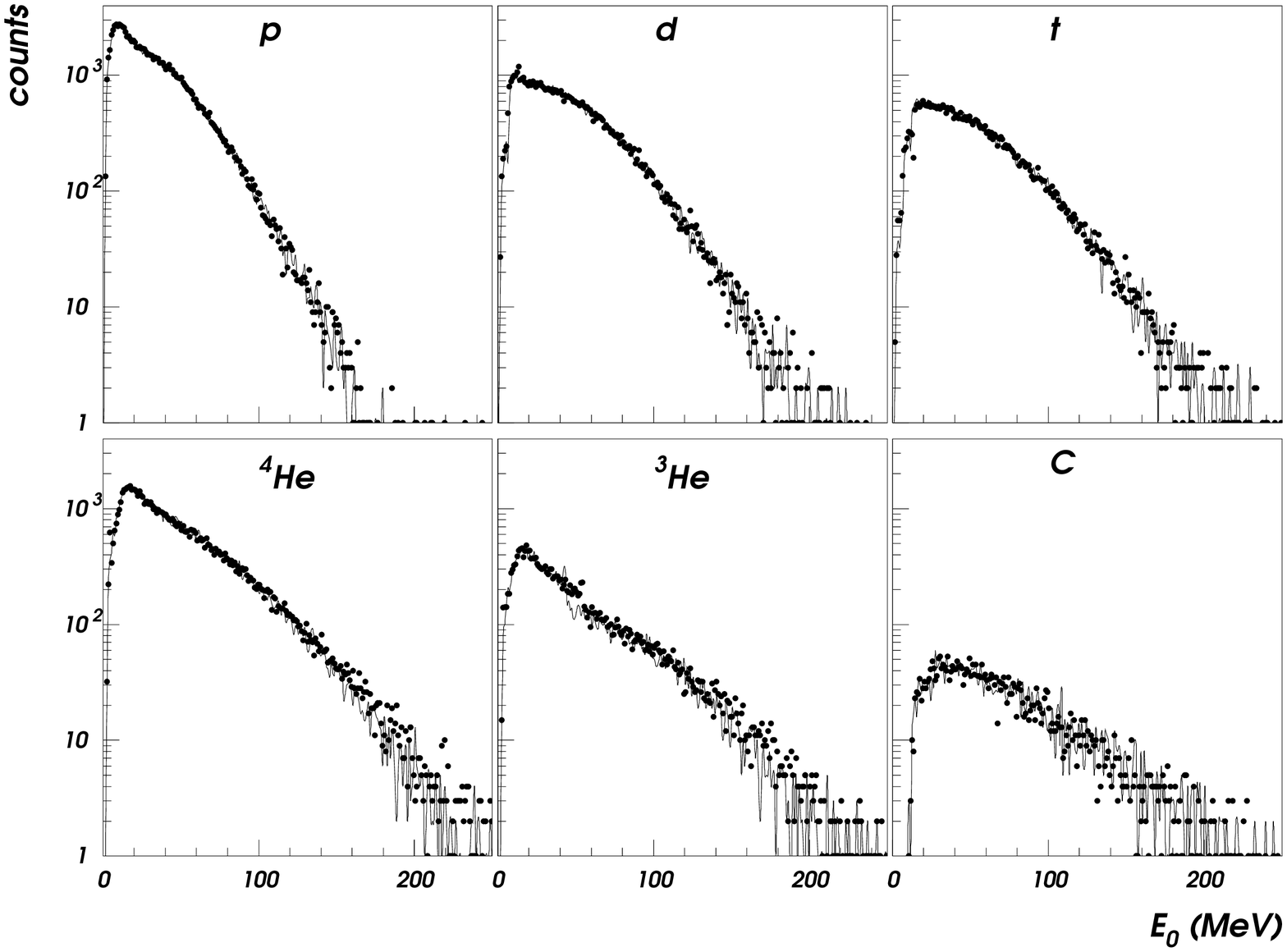}
%\epsfbox{etire.eps}
%\caption{Energy spectra for different particles in module no. 6, ring 10
%(symbols) and reference spectra (line) for the Si(Li) detector on the same
%ring.}
\caption{Energy spectra of light charged particles and fragments in one of the 
CsI(Tl) crystal of ring 10 of INDRA (symbols). They are  compared to the same 
charged
reaction product energy spectra of the reference module (lines) of this ring.}
\label{fig14}
\end{figure}

In the second stage all the other CsI(Tl) scintillators of the ring are
calibrated. The energy spectra of a given Z have to be identical to
the corresponding reference one (the trigger used in the experiments did not
break the axial symmetry of INDRA).
%Light charged particles are well identified
%in each of the $\Delta E_{\mathrm IC} - Q_0$ map.
An energy spectrum, starting from the light response spectrum, is obtained
by stretching it between $E_{\mathrm min}, ~E_{\mathrm max}$ in order
to reproduce the reference spectrum. A $\chi^2$ minimizing procedure
%\cite{riv96}
based on the MINUIT package from CERN library is used. It
provides the parameters $a_{\mathrm 1}$,
 for the considered module. The parameters $a_{\mathrm i}$ (i=1,4) allow
afterwards a complete Z identification in the $\Delta E_{\mathrm IC} - Q_0$ map. 
The energy spectra of the fragments stopped in the respective scintillator are 
determined too \cite{len99}. The good quality of the procedure is illustrated 
in Fig. \ref{fig14}, where energy spectra of different particles stopped in one 
of the scintillators of ring 10 of INDRA are compared to the reference spectra 
of the same ring. 
%%%%%%%%%%%%%% Figura 14
%%%%%%%%%%%%%%%%%%%%%%%%%%%%

%%%%%%%%%%%%%%%%%%%%%%%%%%%%%%%%%%%%%%%%%%%%%%%%%%%%%%%%%%%%%%%%%%%%%%%%%%%%
\subsection{And if the $\delta$ -- rays would be neglected?\label{subsect45}}
We have to stress once again the importance of taking into account the
$\delta$ -- ray effect in the light output, especially for reaction products
heavier than intermediate mass fragments ($Z \geq 15$). By neglecting it,
only the first term in the right side of Eq. (\ref{ecu29}) would
appear. The results remain reasonable for $Z \leq 15$ but with 
values of the fit parameters which change significantly as compared to the
case where the knock-on electrons were taken into account. The 
quality of the fit is drastically degraded for fragments heavier than Z=15.
%, as shown in Fig. \ref{fig9}. 
So are particle
identification (Fig. \ref{fig10}) and energy calibration.
%Fig. \ref{fig11}.
%especially for fragments heavier than intermediate mass fragments.
%%%%%%%%%%%%%%%%%%%%%%%%%%%%%%%%%%%%%%%%%%%%%%%%%%%%%%%%%%%%%%%%%%%%%%%%%%%%%
\section{Conclusions\label{sect5}}
The data obtained with the INDRA array (large range in $Z$ and $E_0$) provided 
a good opportunity to proceed to a more basic study of the light output of 
CsI(Tl) scintillators and to derive suitable calibration and
identification procedures. Starting from the fast and the slow parts of
the light output (fractions of the total light response integrated in the
corresponding time gates), it was possible to rebuild the 
integral of the signal.%, which, for fragments having relatively low average
%stopping power ($AZ^2/E_{\mathrm 0} \geq 0.4$) (most of the data in
%multifragmentation studies) covers 98\% of the estimated true integral, and 
%in the worst case (very energetic protons) more than 90\%. In the latter 
%case, a criterion was sketched to perform the correction.
\par
Under suitable approximations, the expression of the total light output
derived in the preceding paper \cite{par00i} may be analytically
integrated. Even if up to 3\% of the accuracy may be lost, the
fact presents the huge advantage of extremely short computing time. The
derived expressions, easily to handle, were successfully applied for fragment
identification in %$\Delta E_{\mathrm Si, IC} - Light_{\mathrm CsI(Tl)}$
$\Delta E_{\mathrm Si, IC} - Q_0$ maps and
for the energy calibration of the CsI(Tl) scintillators. At forward angles,
where a Si detection layer exists, these applications lead to an important 
reduction of the computing time. At backward angles, where two problems exist:
fragment identification and energy calibration of the CsI(Tl) crystals, 
the above procedure plays an even more important role and corresponds 
to the optimum way we have found to solve these two tasks.
\par
A comparative study of the CsI(Tl) scintillators of INDRA has shown that the
model parameters are meaningful quantities, related to the light collection
and the PMT gain, to the activator and eventually crystal imperfection
concentrations and to the $\delta$ -- ray production energy threshold.
Except the gain parameter, all the others are characteristics of the usual
CsI(Tl) scintillators. Their averages, performed over the 324 CsI(Tl)
crystals of INDRA, allowed to find reliable, recommended parameter values. 
Together with the related total light expression, they constitute good 
implements for energy calibration and heavy ion identification applications in 
heavy ion physics experiments.
%%%%%%%%%%%%%%%%%%%%%%%%%%%%%%%%%%%%%%%%%%%%%%%%%%%%%%%%%%%%%%%%%%%%%%%%%%%%%%


\begin{thebibliography}{99}
\bibitem{pou95}J. Pouthas et al., Nucl. Instr. and Meth. A 357 (1995) 418.
\bibitem{pou96}J. Pouthas et al., Nucl. Instr. and Meth. A 369 (1996) 222.
\bibitem{par00i}M. Parlog et al., Nucl. Instr. and Meth. A, preceding paper
\bibitem{wil66}D. Williams, G. F. Snelling and J. Pickup, Nucl. Instr. and
        Meth. 39 (1966) 141.
\bibitem{str90}D. W. Straecener et al., Nucl. Instr. and Meth. A 294 (1990)
        485.
%\bibitem{bdh}BDH-Merck Ltd, West Quay Rd, Poole, BH15 1HX, England.
%\bibitem{gui82}GUIDE7 (Version 1982), T. Massam, CERN 76-21, Exp. Phys. Div.
%\bibitem{phi}Philips Photonics, F19108 Brive, France.
\bibitem{gui89}D. Guinet et al.,Nucl. Instr. and Meth. A 278 (1989) 614.
\bibitem{laser}Laser Science Inc., Cambridge, MA, USA.
\bibitem{tab96}G. Tabacaru et al., Nucl. Instr. and Meth. A 428 (1999) 379.
\bibitem{hub90}F. Hubert, R. Bimbot, H. Gauvin, At. Data and Nucl. Data
        Tables 46 (1990) 1.
\bibitem{nor}L. C. Northcliffe and R. F. Schillling, Nuclear Data Tables A7
        (1970) 233.
\bibitem{cha98}R. J. Charity, Phys. Rev. C 58 (1998) 1073.
\bibitem{lia74}V. K. Liapidevski et al., PTE 2 (1974) 62.
\bibitem{gra85}H. Grassmann, E. Lorentz and H. G. Moser, Nucl. Instr. and
        Meth. A 228 (1985) 323.
\bibitem{val93}J. D. Valentine, W. M. Moses, S. E. Derenzo, D. K. Wehe and
        G. F. Knoll, Nucl. Instr. and Meth. A 325 (1993) 147.
\bibitem{rob61ii}J. C. Robertson, J. G. Lynch and W. Jack, Proc. Phys. Soc.
        78 (1961) 1188.
\bibitem{rob61i}J. C. Robertson and J. G. Lynch, Proc. Phys. Soc. 77 (1961)
        751.
\bibitem{sto58}R. S. Storey, W. Jack and A. Ward, Proc. Phys. Soc. 72 (1958)
        1.
\bibitem{ben89}F. Benrachi et al., Nucl. Instr. and Meth. A 281 (1989) 137.
\bibitem{kno89}Glenn F. Knoll, Radiation detection and measurements, third
        edition, John Wiley and Sons, Inc. (2000) 229.
%\bibitem{riv96}M. F. Rivet, M. Parlog, E. Plagnol, L. Tassan-Got, Inst.
%       Nucl. Phys. Orsay, INDRA collaboration, IPN-Orsay Report, Sept. 19
%       (1996).
\bibitem{bdh}BDH-Merck Ltd, West Quay Rd, Poole, BH15 1HX, England.
\bibitem{gwi63i}R. Gwin and R. B. Murray, Phys. Rev. 131 (1963) 501.
\bibitem{eby54}F. S. Eby and W. K. Jentschke, Phys. Rev. 96 (1954) 911.
\bibitem{len99}N. Le Neindre, Th\`ese docteur de l'Universit\'e de Caen,
        (1999) LPCC T 99-02.
\end{thebibliography}
\end{document}